\documentclass[10pt, prb, aps, twocolumn, showpacs, citeautoscript, floatfix, reprint, amsmath, amssymb, notitlepage, superscriptaddress]{revtex4-1}

\usepackage{graphicx}
\usepackage{stmaryrd}
\usepackage{rotating}
\usepackage{amsmath}
\usepackage{amsfonts}
\usepackage{amssymb}
\usepackage[countmax]{subfloat}
\usepackage{dcolumn} % align table columns on decimal point
\usepackage{bm} % bold math
\usepackage{color}

\usepackage{float}
\usepackage{latexsym}
\usepackage{wasysym}

\begin{document}

\title{Persistent currents and spin torque caused by percolated quantum spin Hall state}

%\title{Laminar current and spin torque induced by percolated topological edge states}

\author{Antonio Zegarra}

\affiliation{Department of Physics, PUC-Rio, 22451-900 Rio de Janeiro, Brazil}

\author{J. Carlos Egues}
\affiliation{Instituto de F\'{i}sica de S\~ao Carlos, Universidade de S\~ao Paulo, 13560-970 S\~ao Carlos, S\~ao Paulo, Brazil}

\author{Wei Chen}

\affiliation{Department of Physics, PUC-Rio, 22451-900 Rio de Janeiro, Brazil}

\date{\today}

\begin{abstract}

Motivated by recent experiments, we investigate the quantum spin Hall state 
in 2D topological insulator/ferromagnetic metal planar junctions by means of a tight-binding model and linear response theory. We demonstrate that whether the edge state Dirac cone is submerged into the ferromagnetic subbands and the direction of the magnetization dramatically affect (i) how the edge state percolates into the ferromagnet, and (ii) the spin-momentum locking of the edge state. Laminar flows of room temperature persistent charge and spin currents near the interface are uncovered. In addition, the current-induced spin polarization at the edge of the 2D topological insulator is found to be dramatically enhanced near the impurities. The current-induced spin polarization in the ferromagnet is mainly polarized in the out-of-plane direction ${\hat{\bf z}}$, rendering a current-induced spin torque that is predominantly field-like $\propto {\bf S}\times{\hat{\bf z}}$. 

%and the current-induced spin torque is found to be entirely field-like due to the real wave functions of the percolated edge state and the quantum well state of the ferromagnet. On the other hand, a damping-like torque can be induced by impurities when the magnetization has an out-of-plane component.

\end{abstract}

\maketitle

\section{Introduction}

The quantum spin Hall effect (QSHE) represents one of the important properties of two-dimensional (2D) time-reversal (TR) invariant topological insulators (TIs)\cite{Kane05_2,Bernevig06,Bernevig06_2,Konig07,Konig08}. Owing to the existence of edge states, the defining feature of QSHE is the spin current circulating the edge of the system, which motivates a variety of edge state based topological spintronic devices. To exploit the edge spin current, the TI is often made in conjunction with a ferromagnetic metal (FMM), for instance in three-dimensional (3D) TI/FMM heterostructures\cite{Mellnik14,Mahendra18,Shiomi14,RojasSanchez16,Mendes17}, such that the magnetization can be used to affect the edge spin transport or vice versa. On the theoretical side, a significant amount of work has been dedicated to understand the complicated spintronic mechanisms in such a hybrid structure\cite{Yokoyama10,Sakai14,Fischer16,Ndiaye17,Okuma17_2,Ghosh18}. However, to delineate an adequate theoretical description, it is crucial to understand how the QSH state is altered when the TI is made in conjunction with a metallic material, especially given that the boundary condition of the edge state wave function is modified.

Recent experiments have also demonstrated the feasibility of spin to charge interconversion in spintronic devices based on 2D TIs. In particular, the high efficiency of spin pumping and spin-transfer torque observed in monolayer or multilayer transition metal dichalcogenide/ferromagnet (TMD/FMM) heterostructures is exceedingly encouraging\cite{Shao16,Zhang16,Mendes18,Bansal19,Wu20,Fan20}, especially given that these materials can realize the QSHE\cite{Qian14,Ma15,Wu18}. Motivated by these experiments, and also to clarify the role of edge states in these spintronic effects, in this article we investigate the 2D TI/FMM planar junction by means of a lattice model approach. We show that the modification of the QSH state depends significantly on whether the edge state Dirac cone submerges into the FMM subbands, as well as on the direction of the
magnetization. These factors strongly influence the percolation of the edge state into the FMM, as well as the spin-momentum locking in the TI region near the TI/FMM interface. We uncover a number of peculiar dissipationless responses, including the existence of room temperature persistent charge and spin currents that manifest as laminar flows. Moreover, we elaborate that the real wave function of the percolated edge state is crucial to the direction and magnitude of the current-induced spin torque.

The structure of the article is organized in the following manner. In Sec.~II A, we detail the lattice model for the 2D TI/FMM junction, and delineate two different types of band structures and the corresponding percolation of topological edge states in Sec.~II B. We proceed to demonstrate that the asymmetric band structure yields a laminar flow of persistent charge current, as well as elaborating the proximity induced persistent spin current in the system in Sec.~II C. The current-induced spin torque is investigated by means of a linear response theory in Sec.~II D, where we emphasize the field-like nature of the spin torque due to the real wave functions of the percolated edge state and the quantum well state of the FMM. Section III summarizes the results.

\section{BHZ/FMM planar junction}

\subsection{Lattice model}

\begin{figure}[ht]
\begin{center}
\includegraphics[clip=true,width=0.99\columnwidth]{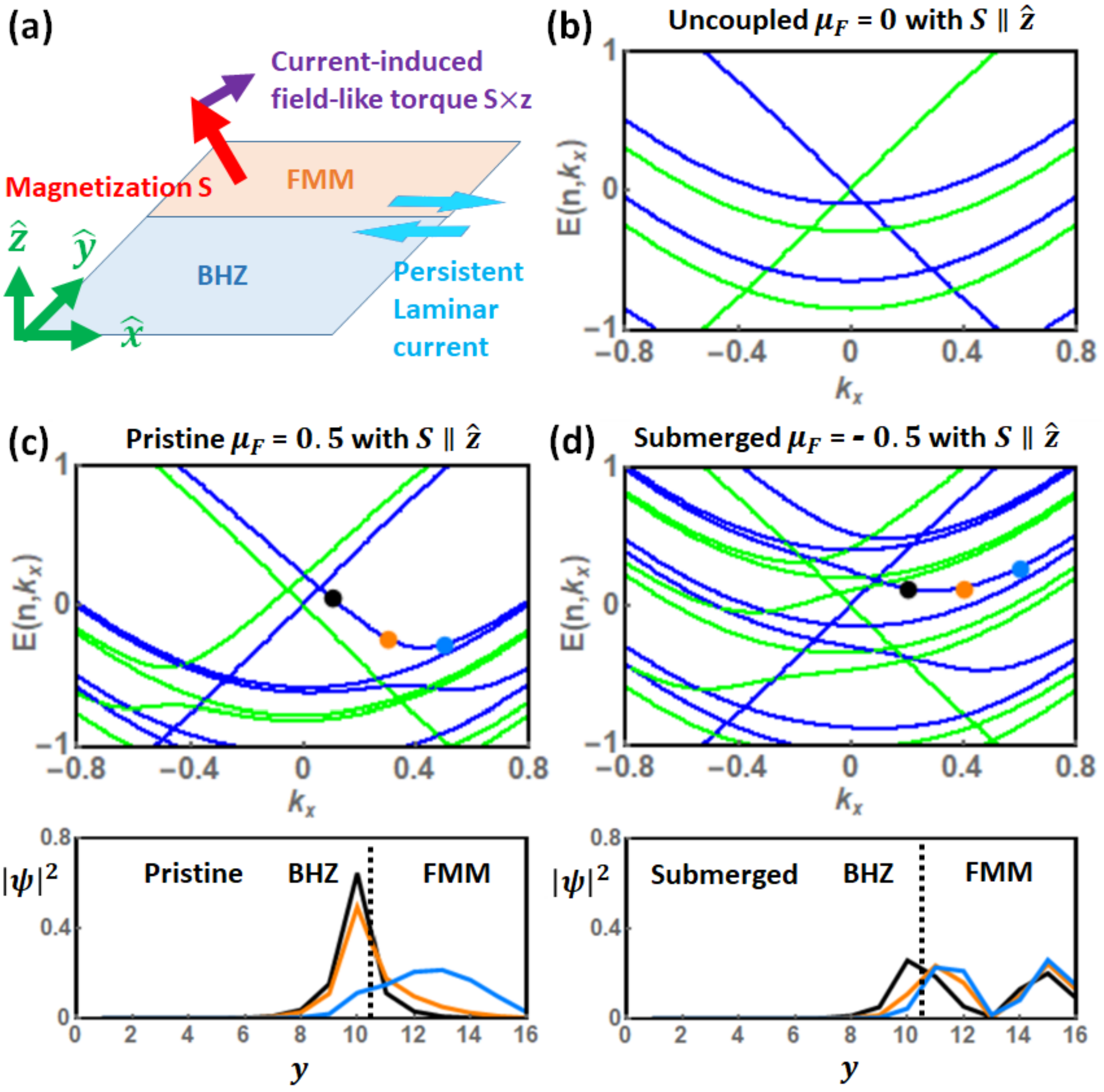}
\caption{(a) Schematics of the lattice model of the BHZ/FMM strip, with PBC along ${\hat{\bf x}}$ and OBC along ${\hat{\bf y}}$. (b) The low energy spin up (blue) and down (green) band structures when the BHZ and FMM are uncoupled $t_{B}=0$. The magnetization is fixed at ${\bf S}\parallel{\hat{\bf z}}$, with BHZ width $N_{y,TI}=10$ and FMM width $N_{y,FM}=6$. (c) The pristine and (d) the submerged types of band structures for the coupled BHZ/FMM strip at interface hopping $t_{B}=0.8$. The undistorted Dirac cone corresponds to the edge state at the vacuum/BHZ interface $y=1$, whereas the distorted one corresponds to that at the BHZ/FMM interface (dashed line). The bottom panels show the wave function profiles $|\psi|^{2}$ (also equal to $\langle\sigma^{z}\rangle$) of the corresponding states of the same colors on the band structure. } 
\label{fig:Edge_state_BHZFMM}
\end{center}
\end{figure}

To properly address the percolation of the edge state, we employ a tight-binding model approach similar to that used for 3D TIs\cite{Ghosh18}. For concreteness, we consider a strip of 2D Bernevig-Hughes-Zhang (BHZ) model\cite{Bernevig06} of width $N_{y,TI}$ in conjunction with a strip of 2D FMM of width $N_{y,FM}$, as indicated in Fig.~\ref{fig:Edge_state_BHZFMM} (a). Periodic boundary condition (PBC) in the longitudinal ${\bf{\hat x}}$ direction and open boundary condition (OBC) in the transverse direction ${\hat{\bf y}}$ are imposed, i.e., a closed BHZ/FMM ribbon. The BHZ region is composed of the spinful $s$ and $p$ orbitals $\psi=\left(s\uparrow,p\uparrow,s\downarrow,p\downarrow\right)^{T}$, with the Dirac matrices $\gamma_{i}=\left\{\sigma^{z}\otimes s^{x},I\otimes s^{y},I\otimes s^{z},\sigma^{x}\otimes s^{x},\sigma^{y}\otimes s^{x}\right\}$ and the TR operator $T=-i\sigma^{y}\otimes IK$, where $\sigma^{b}$ and $s^{b}$ are Pauli matrices in the spin and orbital spaces, respectively. The model in momentum space reads\cite{Bernevig13} 
\begin{eqnarray}
&&H({\bf k})=\sum_{i=1}^{3}d_{i}({\bf k})\gamma_{i}
=A\sin k_{x}\gamma_{1}+A\sin k_{y}\gamma_{2}
\nonumber \\
&&+\left(M-4B+2B\cos k_{x}+2B\cos k_{y}\right)\gamma_{3}
\nonumber \\
&&=\left(
\begin{array}{cc}
h({\bf k}) & 0 \\
0 & h^{\ast}(-{\bf k})
\end{array}
\right),
\label{BHZ_Dirac_Hamiltonian}
\end{eqnarray}
where $h({\bf k})=\sum_{i=1}^{3}d_{i}({\bf k})\sigma^{i}$, $A$ and $B$ are kinetic parameters, and $M<0$ is the topologically nontrivial phase that hosts the edge state.

We now detail the lattice model of the BHZ/FMM ribbon. Due to the proximity to the TI, the conduction band of the FMM is assumed to be split into $s$-like and $p$-like orbitals, both are subject to the magnetization ${\bf S}$ of the FMM through an exchange coupling. The model is described by
\begin{widetext}
\begin{eqnarray}
H&=&\sum_{i\in TI}\left\{-itc_{is\uparrow}^{\dag}c_{i+ap\uparrow}
-itc_{ip\uparrow}^{\dag}c_{i+as\uparrow}+itc_{is\downarrow}^{\dag}c_{i+ap\downarrow}
+itc_{ip\downarrow}^{\dag}c_{i+as\downarrow}+h.c.\right\}
\nonumber \\
&+&\sum_{i\in TI}\left\{-tc_{is\uparrow}^{\dag}c_{i+bp\uparrow}+tc_{ip\uparrow}^{\dag}c_{i+bs\uparrow}
-tc_{is\downarrow}^{\dag}c_{i+bp\downarrow}+tc_{ip\downarrow}^{\dag}c_{i+bs\downarrow}+h.c.\right\}
\nonumber \\
&+&\sum_{i\in TI}\left(M+4t'-\mu\right)\left\{c_{is\uparrow}^{\dag}c_{is\uparrow}
+c_{is\downarrow}^{\dag}c_{is\downarrow}\right\}
+\sum_{i\in TI}\left(-M-4t'-\mu\right)\left\{c_{ip\uparrow}^{\dag}c_{ip\uparrow}+c_{ip\downarrow}^{\dag}c_{ip\downarrow}\right\}
\nonumber \\
&+&\sum_{i\in TI,\delta}\left(-t'\right)\left\{c_{is\uparrow}^{\dag}c_{i+\delta s\uparrow}-c_{ip\uparrow}^{\dag}c_{i+\delta p\uparrow}
+c_{is\downarrow}^{\dag}c_{i+\delta s\downarrow}-c_{ip\downarrow}^{\dag}c_{i+\delta p\downarrow}+h.c.\right\}-\mu_{F}\sum_{i\in FM,I\sigma}c_{iI\sigma}^{\dag}c_{iI\sigma}
\nonumber \\
&-&t_{F}\sum_{i\in FM,\delta I\sigma}\left\{c_{iI\sigma}^{\dag}c_{i+\delta I\sigma}+c_{i+\delta I\sigma}^{\dag}c_{iI\sigma}\right\}
+\sum_{i\in FM,I\sigma}J_{ex}{\bf S}\cdot c_{iI\alpha}^{\dag}{\boldsymbol\sigma}_{\alpha\beta}c_{iI\beta}
-t_{B}\sum_{i\in BD,I\sigma}\left\{c_{iI\sigma}^{\dag}c_{i+b I\sigma}+c_{i+b I\sigma}^{\dag}c_{iI\sigma}\right\}.\;\;\;
\label{Hamiltonian_BHZFMM}
\end{eqnarray}
\end{widetext}
Here $c_{iI\sigma}$ and $c_{iI\sigma}^{\dag}$ are electron annihilation and creation operators, $I=\left\{s,p\right\}$ is the orbital index, $\delta=\left\{a,b\right\}$ denotes the lattice constant along the two planar directions, $\sigma=\left\{\uparrow,\downarrow\right\}$ is the spin index, $i=\left\{x,y\right\}$ denotes the planar position, and $TI$, $FM$, $BD$ denote the TI region, the FMM region, and the interface sites, respectively. 
In addition, due to the Schottky-Mott rule\cite{Cowley65,Tung14}, i.e., the difference in work functions causes an adjustment of the chemical potentials, the FMM on-site energy $\mu_{F}$ becomes a material-dependent parameter that shifts the FMM bands. The magnetization of the FMM is denoted by the classical vector ${\bf S}=S\left(\sin\theta\cos\varphi,\sin\theta\sin\varphi,\cos\theta\right)$.

To make connection with the real HgTe quantum well parameters, the hopping parameters are chosen as
\begin{eqnarray}
A=2t\approx -3.4{\rm eV}\;,\;\;\;
B=-t'\approx -17eV=10t\;.
\end{eqnarray}
We will treat the hopping $t=A/2=-1.7{\rm eV}\equiv -1$ as the energy unit throughout the article (that is, we take 1.7eV as energy unit). However, we find that in the lattice model, if we take the value $t'=-10t=10$, then the energy spectrum does not clearly show a gap. This is obviously because the higher order term in the $d_{3}$ component. If we simulate it with $4t'-2t'\cos k_{x}a-2t'\cos k_{y}a$ with a large hopping amplitude $t'$, then this term will wash out the bulk gap. This is obviously an artifact of using a lattice model to simulate the continuous HgTe quantum well. For this reason we reduce the $t'=-10 t=10$ to $t'\approx -t=1$ in our lattice model in order to maintain the bulk gap and demonstrate the edge state.

The other approximation we will use is about the mass term $M$. In reality, $A/M=2t/M$ gives the decay length of the edge state. Because we will simulate the system on a lattice size of the order of $10\times 10$ sites, this means the decay length cannot exceed few lattice sites, otherwise the edge states on the two opposite edges overlap. Therefore for our simulation we choose the mass term to be $M=-1$, which is quite different from real HgTe quantum wells. The calculations of persistent currents and the magnetoelectric susceptibility (see below) are performed at a scale of the order of room temperature $k_{B}T=0.03$. Finally, the interface hopping, assumed to be between the same orbital and spin species, is fixed at $t_{B}=0.8$ for concreteness. In summary, we use the parameters (in units of $|t|=1.7$eV)
\begin{eqnarray}
&&-t=t'=-M=t_{F}=1\;,\;\;\;\mu=0\;,\;\;\;\mu_{F}=0.5\,({\rm pristine})
\nonumber \\
&&\mu_{F}=-0.5\,({\rm submerged})\;,\;\;\;t_{B}=0.8\;,\;\;\;J_{ex}=0.1\;,
\nonumber \\
&&k_{B}T=0.03\;,
\end{eqnarray}
We emphasize that the statements made in the present work are fairly robust against changing these parameters.

\begin{figure}[ht]
\begin{center}
\includegraphics[clip=true,width=0.99\columnwidth]{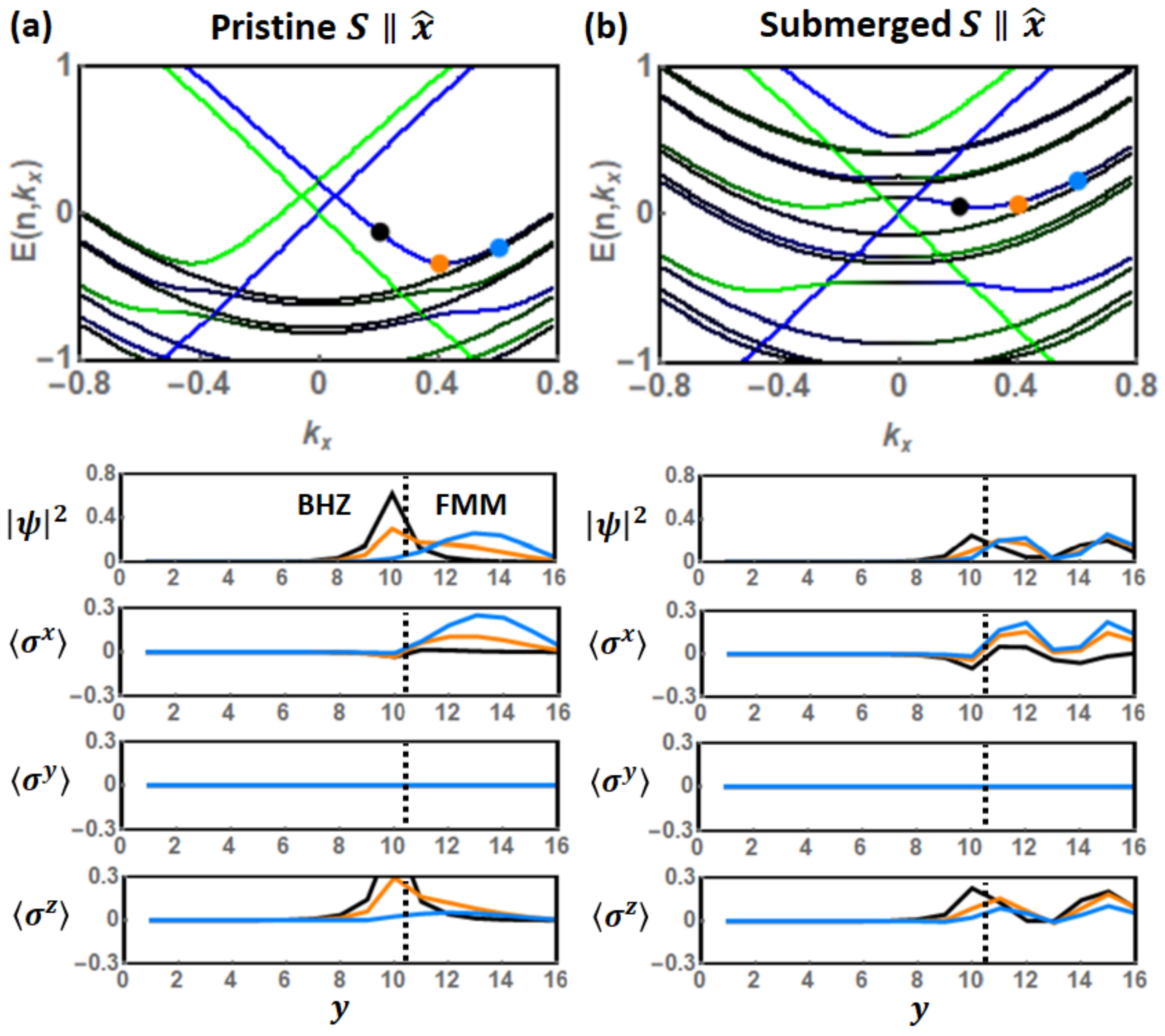}
\caption{The band structures of the BHZ/FMM ribbon with magnetization along ${\bf S}\parallel{\hat{\bf x}}$, for (a) the pristine $\mu_{F}=0.5$ and (b) the submerged $\mu_{F}=-0.5$ type. The green and blue colors indicate the spin up and down polarizations, and the black color parts are unpolarized. The edge state wave function and spin polarization of the corresponding states in the dispersion are shown in the same colors. } 
\label{fig:dispersion_edge_SxSy}
\end{center}
\end{figure}

\subsection{Band structure and percolation of the edge state}

The band structure $E(n,k_{x})$ can be obtained from Eq.~(\ref{Hamiltonian_BHZFMM}) by a partial Fourier transform $c_{iI\sigma}=c_{xyI\sigma}=\sum_{k_{x}}e^{ik_{x}x}c_{k_{x}yI\sigma}$, where $c_{iI\eta}$ is the electron annihilation operator of orbital $I=\left\{s,p\right\}$ and spin $\sigma=\left\{\uparrow,\downarrow\right\}$ at site $i=\left\{x,y\right\}$. For comparison, in Fig.~\ref{fig:Edge_state_BHZFMM} (b) we show the band structure when the BHZ and the FMM are uncoupled $t_{B}=0$, in which the edge state Dirac cone and the quadratic FMM bands are clearly distinguishable. The FMM wave functions are confined quantum well states since the FMM is sandwiched between the TI and the vacuum. Figure ~\ref{fig:Edge_state_BHZFMM} (c) shows what we call the pristine type of band structure for the coupled BHZ/FMM strip simulated by $\mu_{F}=0.5$ and interface hopping $t_{B}=0.8$, and the corresponding percolations of the edge state, with magnetization ${\bf S}\parallel{\hat{\bf z}}$ pointing along the spin polarization of the edge state. The Dirac cone remains gapless, and at larger momenta gradually merges with the FMM subbands of the same spin polarization. Going from small to large momentum, the edge state wave function $|\psi|^{2}=\sum_{I\sigma}|\psi_{I\sigma}|^{2}$ gradually evolves from that highly localized at the edge to a profile that merges with the FMM quantum well state of the first harmonic. Because the edge state Dirac cone is still identifiable, and the feature of wave function merging between the edge state and the FMM quantum well state, we call this state the percolated QSH state.

The other type of band structure simulated by $\mu_{F}=-0.5$ is what we call the submerged type where the Dirac point overlaps with the FMM subbands, as shown in Fig.~\ref{fig:Edge_state_BHZFMM} (d). In this case the Dirac cone at the BHZ/FMM interface is very much distorted and becomes highly intertwined with FMM subbands. Tracking the states originating from the Dirac cone shows that the Dirac cone splits into different branches, each branch hybridizes with the FMM quantum well state of a different harmonic, such as the second harmonic shown by the $|\psi|^{2}$ in Fig.~\ref{fig:Edge_state_BHZFMM} (d). The percolation in both situations also increase with the interface hopping $t_{B}$, as expected (not shown). Although the highly intertwined Dirac cone and FMM subbands make it rather ambiguous to identify edge states at the BHZ/FMM interface, one should keep in mind that the edge states at the other edge $y=1$ remains unaltered, and hence we still regard this submerged situation a QSH state. Finally, whether the Dirac point submerges into the FMM subbands also depends on the number of the FMM subbands, which is given by the width $N_{y,FM}$ of the FMM. For either the pristine or submerged situation, the edge state at the vacuum/BHZ interface at $y=1$ is unaffected by the contact to the FMM at $y=N_{y,TI}$ interface, and the Dirac cone therein remains undistorted.

The simulations for the two other magnetization directions ${\bf S}\parallel{\hat{\bf x}}$ and ${\bf S}\parallel{\hat{\bf y}}$ that are orthogonal to the spin polarization of the edge state are shown in Fig.~\ref{fig:dispersion_edge_SxSy}. The results reveal that the merger between edge states and quantum well states induces a small spin polarization along ${\bf S}$ for the edge states in the BHZ region near the interface (a small $\langle\sigma^{x}\rangle$ near $y\apprle 10$ in Fig.~\ref{fig:dispersion_edge_SxSy}). This indicates that the spin polarization in the BHZ region is no longer perfectly along ${\hat{\bf z}}$, hence the spin-momentum locking is altered by the presence of the magnetization. Likewisely, the percolated edge state in the FMM region is polarized in the plane spanned by ${\bf S}$ and ${\hat{\bf z}}$, instead of entirely along ${\bf S}$, indicating the spin polarization is also distorted in this region. For instance, for either the pristine or the submerged type of band structure, the spin polarization in the ${\bf S}\parallel{\hat{\bf x}}$ case entirely lies in the $xz$-plane. As we shall see in Sec.~\ref{sec:current_induced_spin_torque}, such a peculiar spin texture eventually yields a current-induced spin torque that is entirely field-like.

%a remarkable feature, namely despite the percolation and merging with the quantum well states, the spin polarization of the edge state wave function in the BHZ region remains quantized along ${\hat{\bf z}}$ regardless of the magnetization direction, meaning that it remains as the symmetry eigenstate of $\sigma^{z}$\cite{Bernevig06,Bernevig13}. In other words, the spin-momentum locking is strictly preserved in the TI region. 

\subsection{Laminar charge and spin currents \label{sec:laminar_currents}}

The dispersion for either the pristine or submerged situation becomes asymmetric between $+k_{x}$ and $-k_{x}$ when the magnetization has a component along the spin polarization of the edge state $S_{z}$, as shown in Fig.~\ref{fig:Edge_state_BHZFMM} (b). This is because such a component makes one branch of the Dirac cone more energetically favorable than the other, similar to what occurs in 2D magnetized Rashba systems\cite{Gambardella11}. Although this asymmetry motivates us to speculate the existence of a persistent charge current\cite{Chen15_Majorana}, one should keep in mind that an asymmetric dispersion does not yield a nonzero net current. This can be seen by noticing that the expectation value of the velocity operator $v_{x}$ for the eigenstate $|u_{n,k_{x}}\rangle$ is simply the group velocity\cite{Nagaosa08}
\begin{eqnarray}
\langle u_{n,k_{x}}|v_{x}|u_{n,k_{x}}\rangle=\langle u_{n,k_{x}}|\frac{1}{\hbar}\frac{\partial H}{\partial k_{x}}|u_{n,k_{x}}\rangle=\frac{\partial E(n,k_{x})}{\hbar\partial k_{x}}.
\nonumber \\
\end{eqnarray}
The expectation value of the current operator integrated over momentum vanishes identically
\begin{eqnarray}
\langle v_{x}\rangle=\sum_{n}\int_{-
\pi}^{\pi}\frac{dk_{x}}{2\pi}\frac{\partial E(n,k_{x})}{\hbar\partial k_{x}}f(E(n,k_{x}))=0\;,
\label{ground_state_current}
\end{eqnarray}
where $f(E(n,k_{x}))=1/\left(e^{E(n,k_{x})/k_{B}T}+1\right)$ is the Fermi function, and hence there is no net current.

\begin{figure}[ht]
\begin{center}
\includegraphics[clip=true,width=0.99\columnwidth]{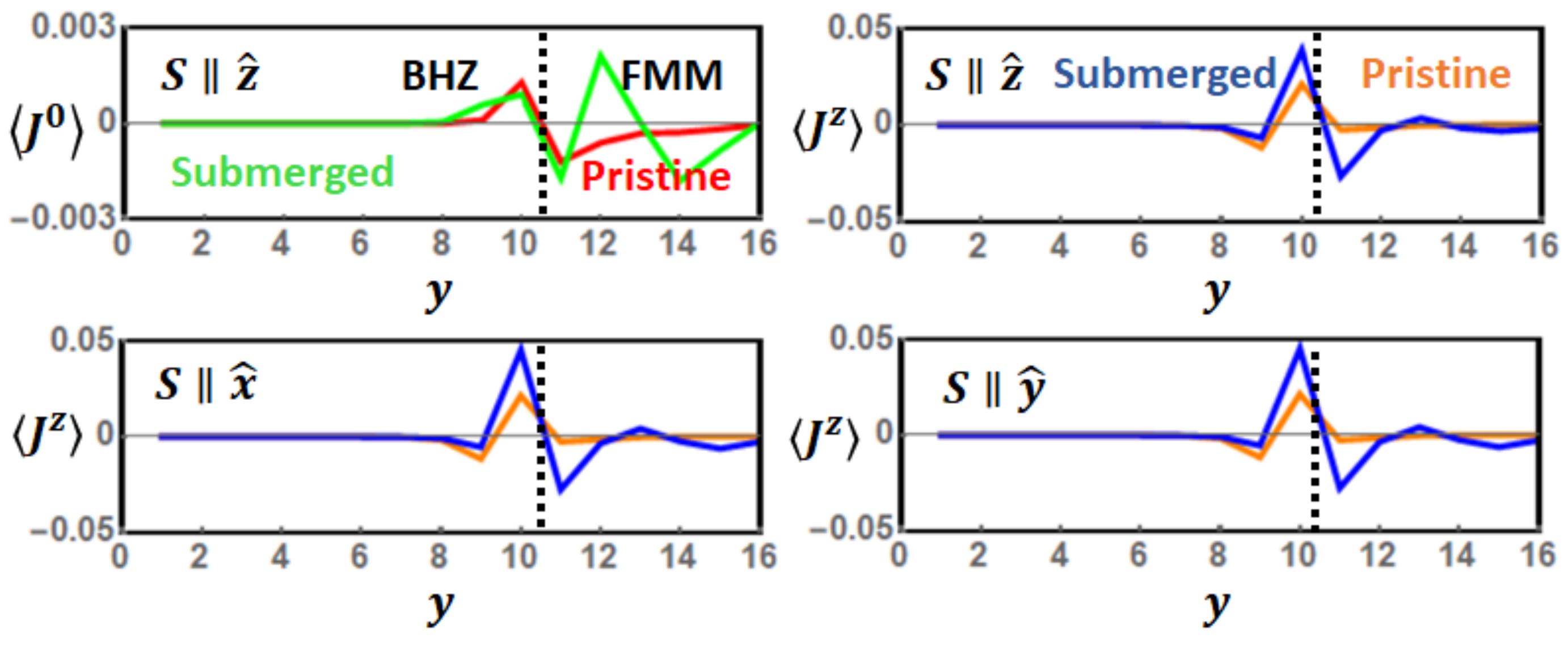}
\caption{ The laminar charge current $\langle J^{0}({y})\rangle$ and spin current $\langle J^{z}({y})\rangle$ as a function of transverse coordinate $y$, at different magnetization directions ${\bf S}\parallel\left\{{\hat{\bf x}},{\hat{\bf y}},{\hat{\bf z}}\right\}$ and for either the pristine or submerged type of band structure. The charge current is nonzero only when the magnetization has a ${\hat{\bf z}}$ component, and both the spin and charge currents vanish if the BHZ and FMM are decoupled $t_{B}=0$.  } 
\label{fig:persistent_current}
\end{center}
\end{figure}

However, the local current is nonzero. This can be seen by evaluating the charge and spin currents directly from the lattice model according to the following procedure. Firstly, the BHZ model does not commute with $\sigma^{x}$ and $\sigma^{y}$, so we only investigate the longitudinal charge current and the spin current polarized along $\sigma^{z}$, and consider the charge/spin polarization operator
\begin{eqnarray}
P^{a}=\sum_{iI\eta\lambda}x_{i}c_{iI\eta}^{\dag}\sigma^{a}_{\eta\lambda}c_{iI\lambda}
\equiv\sum_{I\eta\lambda}P^{a}_{I\eta\lambda}\;,
\label{polarization_operator}
\end{eqnarray}
where $x_{i}$ is the longitudinal coordinate of site $i$, and $\sigma^{a}=\left\{\sigma^{0},\sigma^{z}\right\}=\left\{I,\sigma^{z}\right\}$. The current operators are then $J^{a}={\dot P^{a}}=\frac{i}{\hbar}\left[H,P^{a}\right]$, as calculated explicitly in Appendix \ref{app:current_operators}. The ground state expectation value of the current operator gives the local current
\begin{eqnarray}
\langle J^{a}\rangle=\sum_{n}\langle n|J^{a}|n\rangle f(E_{n}),
\end{eqnarray}
where $|n\rangle$ is the eigenstate with eigenenergy $E_{n}$ of the BHZ/FMM lattice model, and one may separate $\langle J^{a}\rangle$ into contributions from each bond connecting site $i$ and $i+a$ to investigate the local current.

The longitudinal charge current as a function of transverse coordinate $\langle J^{0}(y)\rangle$ is shown in Fig.~\ref{fig:persistent_current}, which features a laminar current whose direction of flow depends on $y$. The net current vanishes up to numerical precision, in accordance with Eq.~(\ref{ground_state_current}). The local charge current is finite only when the magnetization has an out-of-plane component $S_{z}$, a feature inherited from the asymmetric band structure. Moreover, the current is nonzero only when the BHZ and FMM are coupled $t_{B}\neq 0$, so it is entirely proximity induced. A close inspection reveals that both the charge and spin currents arise from contributions not only from the edge states, but from all the subbands. This makes the currents easily persist up to room temperature, which is an advantage over that induced at the topological superconductor/FMM interface\cite{Brydon13,Schnyder13}. For our choice of parameters, the magnitude of the current is of the order of $\langle J^{0}(y)\rangle\sim 10^{-3}et/\hbar\sim 10^{-7}$A, and the flow direction alternates between $+{\hat{\bf x}}$ and $-{\hat{\bf x}}$ at the length scale of lattice constant $\sim$nm. The Ampere's circuital law $B=\mu_{0}\langle J^{0}(y)\rangle/2\pi r$ then indicates that at a distance $r\sim$nm above the surface, the laminar current produces a magnetic field $\sim 1$Oe that points along ${\hat{\bf y}}$ and alternates at the length scale of nm. Thus although the laminar current is not expected to manifest in the transport properties, the alternating magnetic field it produces should in principle be measurable.

\begin{figure}[ht]
\begin{center}
\includegraphics[clip=true,width=0.99\columnwidth]{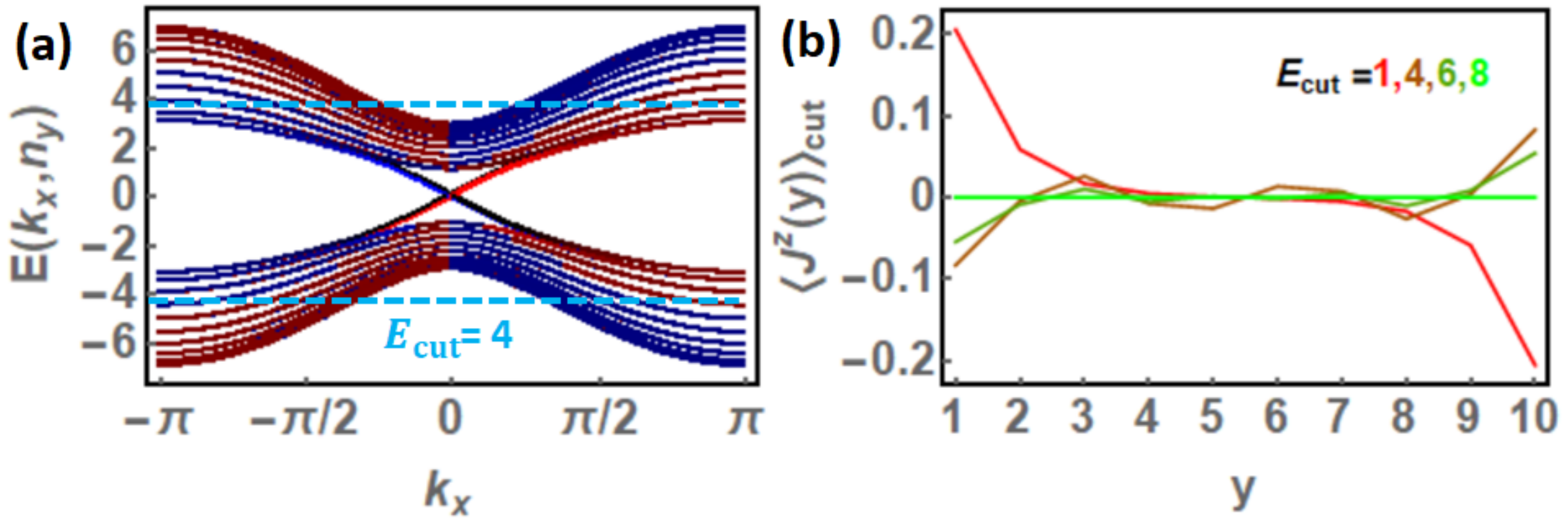}
\caption{ (a) The dispersion of BHZ ribbon of width $N_{y}=10$, with Dirac point located at chemical potential (zero energy). The red and blue color indicate the spin up and down component of the wave function closer to the $y=1$ edge $\tilde{m}_{k_{x},n_{y}}$ defined in Eq.~(\ref{BHZ_m_y1_edge}). (b) the resulting persistent spin current versus transverse coordinate $\langle J^{z}(y)\rangle_{cut}$ calculated by summing only the states within an energy window $|E(n,k_{x})|<E_{cut}$. The $E_{cut}=1$ case includes only the Dirac cone contribution, whereas the $E_{cut}=8$ case sums over the entire band structure. The $E_{cut}=4$ case is shown schematically by the two dashed lines in (a), between which the states are summed and weighted by the Fermi function. } 
\label{fig:BHZ_dispersion_spin_current}
\end{center}
\end{figure}

Concerning the spin current, we first remark that the BHZ model alone does not produce a net edge spin current if the Dirac point locates at the chemical potential. This is because the spin current caused by the edge state is canceled out by the contribution from the BHZ valence bands that are also spin polarized. To elaborate this statement, in Fig.~\ref{fig:BHZ_dispersion_spin_current} (a) we show the dispersion of the BHZ ribbon, with the red and blue colors indicating the spin polarization of the eigenstates $|k_{x},n_{y}\rangle$ near the $y=1$ edge
\begin{eqnarray}
\tilde{m}_{k_{x},n_{y}}^{z}=\sum_{1\leq y\leq N_{y}/2}\sigma_{k_{x},n_{y}}^{z}.
\label{BHZ_m_y1_edge}
\end{eqnarray}
The Dirac cone has the spin up propagating with positive group velocity and spin down with negative group velocity at the $y=1$ edge, as expected. In addition to this, one sees that the valence bands are also spin polarized (so are the conduction bands, but they are not important due to the Fermi distribution). Moreover, at least some parts of the valence bands have spin up but negative group velocity (red color and negative slope in Fig.~\ref{fig:BHZ_dispersion_spin_current} (a)), meaning that these states produce a spin current {\it against} that produced by the edge states. The same mechanism also happens at the other edge $y=10$.

%As shown in Fig.~\ref{fig:BHZ_dispersion_spin_current} (a), the dispersion of the BHZ ribbon displays the well known edge state Dirac cone with spin polarized along ${\hat{\bf z}}$, and the Dirac point is located at zero energy. The persistent charge current $\langle J^{0}(y)\rangle$ is zero everywhere. Quite surprisingly, the spin current $\langle J^{z}(y)\rangle$ at either the edge $y=1$ or the edge $y=N_{y,TI}$ also vanishes even though the Dirac cones consist of counter propagating spins at both edges. 

To quantify the contribution to the edge current from the Dirac cone and that from the valence bands, we calculate the spin current in the lattice model by summing the states within an energy window $E_{cut}$ around the chemical potential
\begin{eqnarray}
\langle J^{z}\rangle_{cut}=\sum_{n}\langle n|J^{z}|n\rangle f(E_{n})\theta(E_{cut}-|E_{n}|)\;,
\end{eqnarray}
where $\theta(E_{cut}-|E_{n}|)$ is the step function. As shown in Fig.~\ref{fig:BHZ_dispersion_spin_current} (b), the $E_{cut}=1$ case that includes only the Dirac cone contribution has a finite spin current, but the $E_{cut}=8$ case that sums over the entire band structure gives a zero spin current. In other words, the contribution from the bulk bands cancels out that from the edge state Dirac cone to yield a zero spin current. A finite spin current occurs only when the Dirac point is shifted away from the chemical potential, or in a certain experiment that can measure the equilibrium spin current contributed only within an energy window near the chemical potential\cite{Chen20_absence_edge_current}.

On the other hand, when the BHZ model is made in conjunction with an FMM, a persistent spin current is produced for both the pristine and the submerged cases, and is a laminar flow that percolates into the FMM, as shown in Fig.~\ref{fig:persistent_current}. Such a laminar spin current appears regardless of the direction of the magnetization and the energy of the Dirac point. The magnitude of the profile of the spin current only differ by about $20\%$ at different magnetization directions, as can be seen by comparing the $\langle J^{z}\rangle$ as a function of $y$ at ${\bf S}\parallel{\hat{\bf x}}$, ${\bf S}\parallel{\hat{\bf y}}$, and ${\bf S}\parallel{\hat{\bf z}}$ in Fig.~\ref{fig:persistent_current}.

\subsection{Current-induced spin torque \label{sec:current_induced_spin_torque}}

The components $b=\left\{x,y,z\right\}$ of the spin polarization induced by a longitudinal electric field $E(i,t){\hat{\bf x}}$
\begin{eqnarray}
\sigma^{b}(i,t)=\chi^{b}(i,\omega)E(i,t)\;.
\label{sigma_chi_E}
\end{eqnarray} 
in our lattice model can be formulated within a linear response theory, where the real part of the DC magnetoelectric susceptibility 
is calculated by\cite{Chen09_resistivity_upturn,Takigawa02}
\begin{eqnarray}
&&\lim_{\omega\rightarrow 0}{\rm Re}\chi^{b}(i,\omega)
\nonumber \\
&&=-\sum_{j}\sum_{m,n}\langle n|\sigma^{b}(i)|m\rangle\langle m|J^{0}(j)|n\rangle\tilde{F}(E_{n},E_{m})\;,
\label{chi_Fnm}
\end{eqnarray}
as detailed in Appendix \ref{app:linear_response}.
The function $\tilde{F}(E_{n},E_{m})$ is highly peaked at $E_{n}\approx E_{m}\approx 0$, meaning that the states at the Fermi surface contribute the most to the response, as expected, which include both the Dirac cone-like bands and the FMM-like subbands according to Fig.~\ref{fig:Edge_state_BHZFMM} (c) and (d). We focus on the DC magnetoelectric susceptibility $\lim_{\omega\rightarrow 0}{\rm Re}\chi^{b}(i,\omega)\equiv\chi^{b}(y)$ as a function of transverse coordinate $y$. We also remark that a recent work in magnetized BHZ model suggests that a damping-like spin torque can be induced by impurities\cite{Ghosh17}. This feature is analogous to the spin mixing enhanced by disorder-induced spin-dependent scattering originally uncovered in metallic spin valves and domain walls\cite{Zhang02,Zhang04}. Motivated by these earlier works, and also for the sake of removing the numerical ambiguities detailed in Appendix \ref{app:linear_response}, we add random impurities into the BHZ/FMM junction
\begin{eqnarray}
H_{imp}=U_{imp}\sum_{i\in imp,I\sigma}c_{iI\sigma}^{\dag}c_{iI\sigma},
\end{eqnarray}
where $i\in imp$ denotes the impurity sites. We choose a relatively large impurity potential $U_{imp}=4$ and density $n_{imp}=10\%$ in an attempt to draw relevance to TMD-based 2D TIs, where a significant amount of defects, such as missing sulfur atoms, are known to be a realistic issue\cite{Liu13,Hong15,Lin16,Vansco16}. The magnetoelectric susceptibility $\chi^{b}$ is then calculated by Eq.~(\ref{chi_Fnm}) using the lattice eigenstates $|n\rangle$.

\begin{figure}[ht]
\begin{center}
\includegraphics[clip=true,width=0.99\columnwidth]{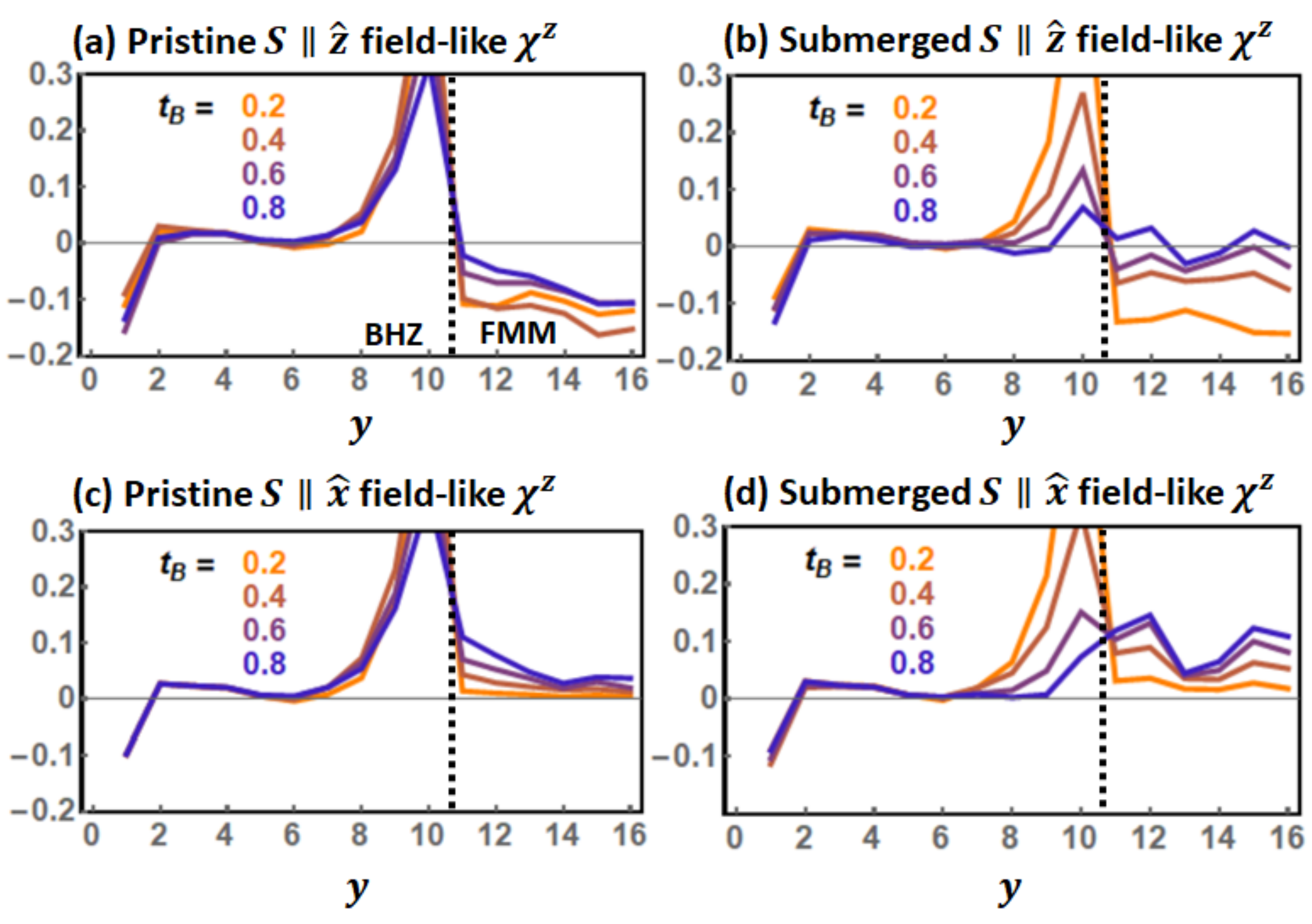}
\caption{The field-like component of the magnetoelectric susceptibility $\chi^{z}(y)$ in the BHZ/FMM junction as a function of transverse coordinate $y$  and at several values of interface hopping $t_{B}$, plotted for different types of band structures and magnetization directions: (a) Pristine ${\bf S}\parallel{\hat{\bf z}}$, (b) submerged ${\bf S}\parallel{\hat{\bf z}}$, (c) pristine ${\bf S}\parallel{\hat{\bf x}}$, and (d) submerged ${\bf S}\parallel{\hat{\bf x}}$. The other two components remain zero $\chi^{x}=\chi^{y}=0$. The $y\leq 10$ is the BHZ region and $y>10$ the FMM region. The negative value near the free edge $y=1$ delineates the Edelstein effect of the BHZ model alone.} 
\label{fig:spin_accumulation}
\end{center}
\end{figure}

%The results of $10\%$ impurities of strength $U_{imp}=4$ are shown in Fig.~\ref{fig:susceptibility_impurity_result}, where the local magnetoelectric susceptibility averaged over longitudinal direction $\sum_{x}\chi^{b}(x,y)/N_{x}$ for an impurity configuration is presented for both types of band structures.

The result for the magnetization directions ${\bf S}\parallel{\hat{\bf z}}$ and ${\bf S}\parallel{\hat{\bf x}}$ is shown in Fig.~\ref{fig:spin_accumulation}, where the magnetoelectric susceptibility averaged over the longitudinal direction $\chi^{z}(y)\equiv\sum_{x=1}^{N_{x}}\chi^{z}(x,y)/N_{x}$ for a specific impurity configuration is presented. The nonzero $\chi^{z}(y)$ near the free edge $y=1$ delineates the Edelstein effect of the BHZ model alone, i.e., current induced spin polarization caused by the edge state, analogous to that occurs in 3D TIs\cite{Tian15,Kondou16,Liu18,Dankert18,Chen20_TI_Edelstein}. At mean free time $\tau\sim 10^{-14}$s and a typical experimental electric field strength $E\sim 10^{4}$kgm/Cs$^{2}$, the induced spin polarization at the free edge is of the order of $10^{-7}$ (in units of $\mu_{B}$). In contrast, at the $y=10$ edge where the BHZ model is made in contact with the FMM, the magnitude of $\chi^{z}(y)$ is enhanced at small interface hopping $t_{B}=0.2$ but decreases at larger $t_{B}$. The spatial profile of $\chi^{z}(y)$ extends into the FMM for both the pristine and the submerged situations, and changes with $t_{B}$ in a rather complicated manner. The band structures in Fig.~\ref{fig:Edge_state_BHZFMM} (c) and (d) naturally explain this enhancement of $\chi^{z}(y)$ due to interface hopping: Compared to an isolated BHZ model, the BHZ/FMM junction has many additional FMM states at the chemical potential ($|n\rangle$ and $|m\rangle$ in Eq.~(\ref{chi_Fnm})) that participate in the particle-hole excitation process of the magnetoelectric response. Moreover, the FMM wave functions and the edge state wave functions have a significant overlap due to percolation of the edge state, yielding nonzero matrix elements $\langle n|{\hat O}|m\rangle$ in Eq.~(\ref{chi_Fnm}). Notice that an isolated FMM does not exhibit Edelstein effect, so the nonzero $\chi^{z}(y)$ in the FMM region $y\in FM$ entirely originates from the proximity to the BHZ model.

%as can be seen by comparing $\chi^{b}(y)$ at different values of interface hopping $t_{B}$.

The average magnetoelectric susceptibility in the FMM region $\chi^{b}_{F}\equiv\sum_{y\in FM}\chi^{b}(y)/N_{y,FM}$ is what yields the spin torque on the magnetization ${\bf S}$. Since the current-induced spin polarization is polarized along ${\hat{\bf z}}$ for an isolated BHZ model, it is customary to define the field-like torque in the FMM to be along ${\hat{\bf S}}\times{\hat{\bf z}}$ and the damping-like torque to be along ${\hat{\bf S}}\times({\hat{\bf S}}\times{\hat{\bf z}})$, as in the usual metallic thin film spin-transfer torque (STT) devices. We find that if the magnetization lies in the $xy$-plane or entirely points along ${\hat{\bf z}}$, then out of the three components ${\chi}_{F}^{b}=\left\{\chi_{F}^{x},\chi_{F}^{y},\chi_{F}^{z}\right\}$ only $\chi_{F}^{z}$ is nonzero. This indicates that the spin torque is entirely field-like if the magnetization lies in the $xy$-plane, and there is no torque if the magnetization points out-of-plane ${\bf S}\parallel{\hat{\bf z}}$. For magnetization along other directions, a small damping-like component develops (note that our calculation neglects other complications such as spin-orbit torque\cite{Manchon08,Manchon09} and spin relaxation). This is very different from the STT in usual metallic heterostructures\cite{Berger96,Slonczewski96} or that induced by the spin Hall effect\cite{Chen15_spin_transfer_torque,Sakanashi18}, where the plane wave states usually contribute to both field-like and damping-like torque at any magnetization direction.

%{\cblue (1) Maybe I should figure out why the torque is enhanced near the impurity site. }

\begin{figure}[ht]
\begin{center}
\includegraphics[clip=true,width=0.99\columnwidth]{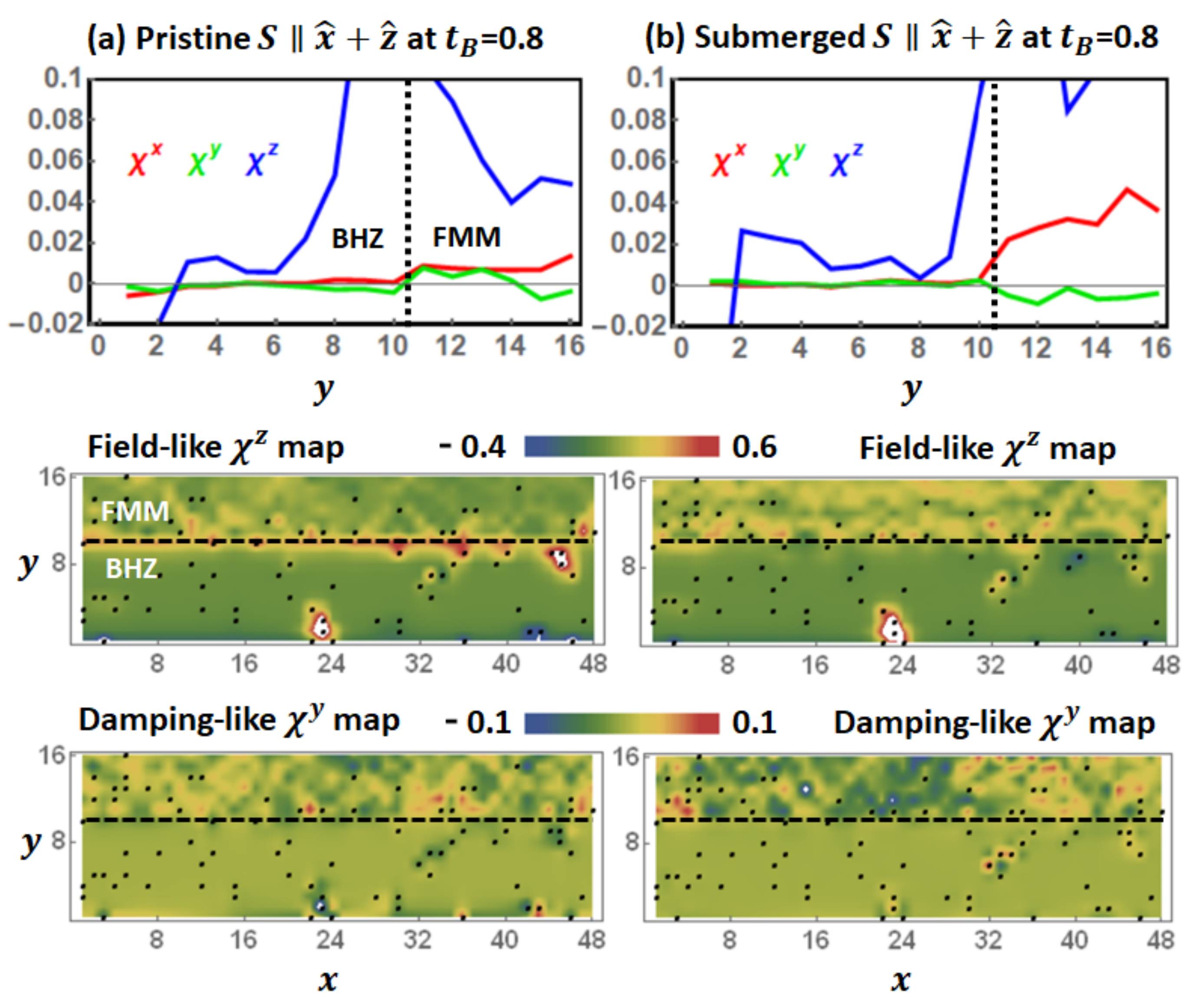}
\caption{Magnetoelectric susceptibility $\chi^{b}$ as magnetization points at ${\bf S}\parallel{\hat{\bf x}}+{\hat{\bf z}}$ averaged over longitudinal position $x$ at a specific impurity configuration, for both the (a) pristine and (b) submerged types of band structures. The spatial distribution of the field-like $\chi^{z}$ damping-like $\chi^{y}$ components are shown in lower panels, where the black dots label the impurity positions. } 
\label{fig:susceptibility_impurity_result}
\end{center}
\end{figure}

Figure \ref{fig:susceptibility_impurity_result} shows the result for the magnetization direction ${\bf S}\parallel{\hat{\bf x}}+{\hat{\bf z}}$, which has both the field-like $\chi^{z}$ and damping-like $\chi^{y}$ components. We find that the damping-like component is generally one order of magnitude smaller than the field-like component. A closer investigation shows that near the two edges of the BHZ region, both $\chi^{z}$ and $\chi^{y}$ are locally enhanced by the impurities, as can be seen from the local $\left\{\chi^{y},\chi^{z}\right\}$ map in Fig.~\ref{fig:susceptibility_impurity_result}. On the other hand, in the FMM region, the magnitude of both components does not seem to correlate with impurity positions. These features are true for both the pristine and the submerged types of band structures. At a typical external electric current $j_{c}\sim 10^{11}$A/m$^{2}$, the spin polarization obtained from Eq.~(\ref{sigma_chi_E}) yields a spin torque according to the Landau-Lifshitz dynamics 
\begin{eqnarray}
\frac{d{\bf S}}{dt}=\frac{J_{ex}}{\hbar}\left[\frac{1}{N_{y,FM}}\sum_{i\in FM}{\boldsymbol\sigma}({i})\right]\times {\bf S}\;.
\label{spin_torque_averaged_over_FMM}
\end{eqnarray}
which is basically the numerical values of $\chi^{b}_{F}$ multiplied by GHz, as demonstrated in Appendix \ref{app:linear_response}. The absolute magnitude of the spin torque is fairly consistent with that uncovered experimentally\cite{Shao16,Zhang16,Mendes18,Bansal19,Wu20,Fan20}, although one should keep in mind that our BHZ/FMM side junction is different from the experimental setup where the TMD is usually deposit on top of the FMM thin film.

%This magnitude is close to that observed in 3D TI/FMM bilayer thin films\cite{Mellnik14}, although one should keep in mind that our lattice model is in 2D and with significantly modified parameters that suites the calculation on a small lattice.

\section{Conclusions}

In summary, we address the percolation of QSHE into an adjacent FMM by means of a lattice model. The band structure displays a pristine/submerged dichotomy due to the difference in work functions, which strongly influences the percolation of the edge state. The merger between the edge states and the quantum well states of the FMM modifies the spin momentum-locking near the TI/FMM interface, and also alters the spin polarization in the FMM region. A laminar flow of persistent charge current owing to the asymmetry of the band structure is uncovered, and the edge spin current also turns into a laminar flow that percolate into the FMM. The current-induced spin polarization at the edge of the 2D TI is dramatically enhanced near the impurities. On the other hand, the current-induced spin torque in the FMM is not directly correlated with the impurity positions, and is found to be entirely field-like if the magnetization lies in the $xy$-plane or points at the ${\hat{\bf z}}$ axis. The damping-like component developed at other magnetization directions is generally one order of magnitude smaller than the field-like component. As these results greatly improve our understanding of the role of edge states in these 2D TI-based spintronic effects, it is intriguing to apply our lattice model approach to other situations that are more relevant to experimental setups, such as TI on top of the FMM, which await further investigations.

%due to the real wave function of the percolated edge state and the quantum well state, with a magnitude greatly enhanced by the presence of the FMM subbands. On the other hand, the inclusion of random impurities (or disorder) is shown to induce a damping-like spin torque as long as the magnetization has an out-of-plane component, but its magnitude is much smaller than the field-like component even at the $10\%$ impurities we investigate. Finally, although we restrict our discussions to 2D, we anticipate that these features can also manifest in the contact between TI and FMM in other dimensions, which await to be clarified.

The authors acknowledge the fruitful discussions with M. H. Fischer, A. P. Schnyder, R. Torr\~{a}o, F. Garcia, and V. Carozo, as well as the financial support from FAPESP (grant No 2016/08468-0) and the productivity in research fellowship from CNPq.

\appendix

\section{The current operators \label{app:current_operators}}

The charge and spin current operator of this lattice model can be calculated conveniently in the following manner. Firstly, the system is translationally invariant along ${\hat{\bf x}}$, so we only calculate the currents flowing along ${\hat{\bf x}}$. In the calculation of the current operator from the polarization operator $J^{a}={\dot P^{a}}=\frac{i}{\hbar}\left[H,P^{a}\right]$, one may simplify the tedious commutator $\left[H,P^{a}\right]$ from the following general consideration. Since only hopping terms in Eq.~(\ref{Hamiltonian_BHZFMM}) contribute to the current operator, we focus on these terms that generally take the form
\begin{eqnarray}
H_{L\alpha M\beta}^{\delta}=\sum_{j}T_{L\alpha M\beta}^{\delta}c_{jL\alpha}^{\dag}c_{j+\delta M\beta}+T_{L\alpha M\beta}^{\delta\ast}c_{j+\delta M\beta}^{\dag}c_{jL\alpha}\;,
\nonumber \\
\label{HLalphaMbeta_delta}
\end{eqnarray}
which describes the hopping of electron between site/orbital/spin $jL\alpha$ and $j+\delta M\beta$ along the planar directions $\delta=\left\{a,b\right\}$, with $T_{L\alpha M\beta}^{\delta}$ the hopping amplitude. Using the fact that the hopping part of the total Hamiltonian is the summation of $H_{t}=\sum_{\delta}\sum_{L\alpha M\beta}H_{L\alpha M\beta}^{\delta}$, we obtain that a specific orbital/spin species $I\eta\lambda$ contributes to the charge current ($a=0$) and the spin current ($a=z$) by, following the definition in Eq.~(\ref{polarization_operator}),
\begin{eqnarray}
&&J^{a}_{I\eta\lambda}=\frac{i}{\hbar}\sum_{\delta}\sum_{L\alpha M\beta}\left[H_{L\alpha M\beta}^{\delta},P^{a}_{I\eta\lambda}\right]
\nonumber \\
&&=\frac{i}{\hbar}\sum_{i}
\left\{\sum_{M\beta}\left[-x_{i}T_{I\lambda M\beta}^{a}\right]c_{iI\eta}^{\dag}\sigma^{a}_{\eta\lambda}c_{i+aM\beta}\right.
\nonumber \\
&&+\sum_{L\alpha}\left[(x_{i}+a)T_{L\alpha I\eta}^{a}\right]c_{iL\alpha}^{\dag}\sigma^{a}_{\eta\lambda}c_{i+aI\lambda}
\nonumber \\
&&\;\;\;+\sum_{L\alpha}\left[(-x_{i}-a)T_{L\alpha I\lambda}^{a\ast}\right]c_{i+aI\eta}^{\dag}\sigma^{a}_{\eta\lambda}c_{iL\alpha}
\nonumber \\
&&\;\;\;\left.+\sum_{M\beta}\left[x_{i}T_{I\eta M\beta}^{a\ast}\right]c_{i+aM\beta}^{\dag}\sigma^{a}_{\eta\lambda}c_{iI\lambda}\right\}\;.
\end{eqnarray}
We then put in all the nonzero hopping amplitudes $T_{L\alpha M\beta}^{\delta}$ and $T_{L\alpha M\beta}^{\delta\ast}$ according to Eq.~(\ref{Hamiltonian_BHZFMM}), and sum over all the $I\eta\lambda$ species. The resulting charge current operator reads
\begin{eqnarray}
J^{0}&=&\frac{1}{\hbar}\sum_{i\in TI}\sum_{\sigma}\left\{\eta_{\sigma}t\,c_{is\sigma}^{\dag}c_{i+ap\sigma}
+\eta_{\sigma}t\,c_{i+ap\sigma}^{\dag}c_{is\sigma}\right.
\nonumber \\
&&\left.+\eta_{\sigma}t\,c_{ip\sigma}^{\dag}c_{i+as\sigma}
+\eta_{\sigma}t\,c_{i+as\sigma}^{\dag}c_{ip\sigma}\right\}
\nonumber \\
&+&\frac{1}{\hbar}\sum_{i\in TI}\sum_{\sigma}\left\{-it'\,c_{is\sigma}^{\dag}c_{i+as\sigma}+it'\,c_{i+as\sigma}^{\dag}c_{is\sigma}\right.
\nonumber \\
&&\left.+it'\,c_{ip\sigma}^{\dag}c_{i+ap\sigma}-it'\,c_{i+ap\sigma}^{\dag}c_{ip\sigma}\right\}
\nonumber \\
&+&\frac{1}{\hbar}\sum_{i\in FM}\sum_{\sigma}\left\{-it_{F}\,c_{is\sigma}^{\dag}c_{i+as\sigma}+it_{F}\,c_{i+as\sigma}^{\dag}c_{is\sigma}\right.
\nonumber \\
&&\left.-it_{F}\,c_{ip\sigma}^{\dag}c_{i+ap\sigma}+it_{F}\,c_{i+ap\sigma}^{\dag}c_{ip\sigma}\right\}\;,
\end{eqnarray}
where $\eta_{\uparrow}=1$, $\eta_{\downarrow}=-1$, and $i\in TI$, $i\in FM$ and $i\in BD$ indicate that the sites $i$ and $i+a$ belong to the BHZ model part, the FMM part, and the interface bonds. Likewisely, the operator for spin current polarized along $z$ is 
\begin{eqnarray}
J^{z}&=&\frac{1}{\hbar}\sum_{i\in TI}\sum_{\sigma}\left\{t\,c_{is\sigma}^{\dag}c_{i+ap\sigma}
+t\,c_{i+ap\sigma}^{\dag}c_{is\sigma}\right.
\nonumber \\
&&\left.+t\,c_{ip\sigma}^{\dag}c_{i+as\sigma}
+t\,c_{i+as\sigma}^{\dag}c_{ip\sigma}\right\}
\nonumber \\
&+&\frac{1}{\hbar}\sum_{i\in TI}\sum_{\sigma}\left\{-it'\eta_{\sigma}\,c_{is\sigma}^{\dag}c_{i+as\sigma}+it'\eta_{\sigma}\,c_{i+as\sigma}^{\dag}c_{is\sigma}
\right.
\nonumber \\
&&\left.+it'\eta_{\sigma}\,c_{ip\sigma}^{\dag}c_{i+ap\sigma}-it'\eta_{\sigma}\,c_{i+ap\sigma}^{\dag}c_{ip\sigma}\right\}
\nonumber \\
&+&\frac{1}{\hbar}\sum_{i\in FM}\sum_{\sigma}\left\{-it_{F}\eta_{\sigma}\,c_{is\sigma}^{\dag}c_{i+as\sigma}+it_{F}\eta_{\sigma}\,c_{i+as\sigma}^{\dag}c_{is\sigma}
\right.
\nonumber \\
&&\left.-it_{F}\eta_{\sigma}\,c_{ip\sigma}^{\dag}c_{i+ap\sigma}+it_{F}\eta_{\sigma}\,c_{i+ap\sigma}^{\dag}c_{ip\sigma}\right\}\;,
\end{eqnarray}
which is essentially the same as $J^{0}$ except the spin up and down channel have an additional minus sign difference, as expected.

\section{Linear response theory for the magnetoelectric susceptibility \label{app:linear_response}}

To calculate the spin accumulation induced by a charge current, we employ the linear response theory for the local spin accumulation $\sigma^{b}(i,t)$ in the presence of a perturbation $H'(t')$ in the Hamiltonian
\begin{eqnarray}
\sigma^{b}(i,t)=-i\int_{-\infty}^{t}dt'\langle\left[\sigma^{b}(i,t),H'(t')\right]\rangle\;,
\label{sigma_linear_response}
\end{eqnarray}
where $\sigma^{b}(i,t)=\sum_{I\beta\gamma}c_{iI\beta}^{\dag}(t)\sigma^{b}_{\beta\gamma}c_{iI\gamma}(t)$ is the $b=\left\{x,y,z\right\}$ component of the spin operator at position $i$, and the fermion operators $c_{iI\gamma}(t)$ are defined in the Heisenberg picture. The perturbation comes from the longitudinal component of the vector field $A(j,t')$ that induces the electric field and the electric current, and hence
\begin{eqnarray}
H'(t')=-\sum_{j}J^{0}(j,t')A(j,t')\;,
\end{eqnarray}
where the electric field comes from the time-variation of the vector field $A(i,t)=A(i)e^{-i\omega t}$
\begin{eqnarray}
E=-\partial_{\beta}V-\frac{\partial A}{\partial t}=-\frac{\partial A}{\partial t}=i\omega A\;.
\end{eqnarray}
As a result, the commutator in Eq.~(\ref{sigma_linear_response}) reads 
\begin{eqnarray}
\left[\sigma^{b}(i,t),H'(t')\right]=\frac{i}{\omega}\sum_{j}e^{i\omega(t-t')}E(j,t)\left[\sigma^{b}(i,t),J^{0}(j,t')\right],
\nonumber \\
\end{eqnarray}
since the electric field has a single wave length and frequency $E(i,t)=E^{0}e^{i{\bf q\cdot r}_{i}-i\omega t}$. Consequently, the local spin accumulation in Eq.~(\ref{sigma_linear_response}) becomes 
\begin{eqnarray}
&&\sigma^{b}({\bf r},t)=\sum_{j}\int_{-\infty}^{\infty}dt'e^{i\omega(t-t')}\frac{1}{\omega}\theta(t-t')
\nonumber \\
&&\;\;\;\times\langle\left[\sigma^{b}(i,t),J^{0}(j,t')\right]\rangle E(j,t)
\nonumber \\
&=&\sum_{j}\int_{-\infty}^{\infty}dt'e^{i\omega(t-t')}\frac{i\pi^{b}(i,j,t-t')}{\omega} E(j,t)
\nonumber \\
&=&\sum_{j}\frac{i\pi^{b}(i,j,\omega)}{\omega} E(j,t)\equiv \sum_{j}\chi^{b}(i,j,\omega) E(j,t).
\end{eqnarray}
Here $\chi^{b}(i,j,\omega)$ is the response coefficient for the contribution to the $\sigma^{b}(i,t)$ at site $i$ due to the longitudinal electric field $E(j,t)$ applied at site $j$. We will further assume that the electric field is constant everywhere, i.e., ${\bf q}\rightarrow 0$ such that $E(i,t)=E(j,t)=E^{0}e^{-i\omega t}$. In this case,
\begin{eqnarray}
\sigma^{b}(i,t)=\left\{\sum_{j}\chi^{b}(i,j,\omega)\right\} E(i,t)=\chi^{b}(i,\omega)E(i,t)\;,
\nonumber \\
\end{eqnarray} 
We aim to calculate the real part of the DC magnetoelectric susceptibility 
\begin{eqnarray}
\lim_{\omega\rightarrow 0}{\rm Re}\chi^{b}(i,\omega)=\lim_{\omega\rightarrow 0}{\rm Re}\left\{\frac{i}{\omega}\sum_{j}\pi^{b}(i,j,\omega)\right\}\;,
\label{Rechi_DC_limit}
\end{eqnarray}
Let $|n\rangle$ be the eigenstate with eigenenergy $E_{n}$ after diagonalizing the BHZ/FMM junction described by Eq.~(\ref{Hamiltonian_BHZFMM}), the retarded $\pi^{b}(i,j,\omega)$ operator is given by 
\begin{eqnarray}
\pi^{b}(i,j,\omega)=\sum_{m,n}\langle n|\sigma^{b}(i)|m\rangle\langle m|J^{0}(j)|n\rangle\frac{f(E_{n})-f(E_{m})}{\omega+E_{n}-E_{m}+i\eta}\;,
\nonumber \\
\label{piij_operator}
\end{eqnarray}
where $\eta$ is a small artificial broadening. Using $\eta/(x^{2}+\eta^{2})=\pi\delta_{\eta}(x)$, the limit in Eq.~(\ref{Rechi_DC_limit}) reads
\begin{widetext}
\begin{eqnarray}
&&-\lim_{\omega\rightarrow 0}{\rm Re}\left\{\frac{i}{\omega}\sum_{j}\pi^{b}(i,j,\omega)\right\}=
\lim_{\omega\rightarrow 0}\left\{\sum_{m,n}\langle n|\sigma^{b}(i)|m\rangle\langle m|\sum_{j}J^{0}(j)|n\rangle\frac{f(E_{n})-f(E_{m})}{\omega}
\frac{-\eta}{(\omega+E_{n}-E_{m})^{2}+\eta^{2}}\right\}
\nonumber \\
&&=
\lim_{\omega\rightarrow 0}\left\{\sum_{m,n}\langle n|\sigma^{b}(i)|m\rangle\langle m|\sum_{j}J^{0}(j)|n\rangle\frac{f(E_{n})-f(E_{n}+\omega)}{\omega}(-\pi)\delta_{\eta}(\omega+E_{n}-E_{m})\right\}
\nonumber \\
&&=
\sum_{m,n}\langle n|\sigma^{b}(i)|m\rangle\langle m|\sum_{j}J^{0}(j)|n\rangle\left(\pi\frac{\partial f(E_{n})}{\partial E_{n}}\right)\delta_{\eta}(E_{n}-E_{m})
\equiv
\sum_{m,n}\langle n|\sigma^{b}(i)|m\rangle\langle m|\sum_{j}J^{0}(j)|n\rangle\tilde{F}(E_{n},E_{m})\;,
\end{eqnarray} 
where we have used the fact that ${\rm Re}\left[\langle m|\sum_{j}J^{0}(j)|n\rangle\right]$ is even but ${\rm Im}\left[\langle m|\sum_{j}J^{0}(j)|n\rangle\right]$ is odd in $(n,m)$, ${\rm Re}\left[\langle n|\sigma^{b}(i)|m\rangle\right]$ is even but ${\rm Im}\left[\langle n|\sigma^{b}(i)|m\rangle\right]$ is odd in $(n,m)$, and the real part of $(1/\omega)(f(E_{n})-f(E_{m}))/(\omega+E_{n}-E_{m}+i\eta)$ in the $\eta\rightarrow 0$ and $\omega\rightarrow 0$ limit is even in $(n,m)$ to eliminate several terms in the $\sum_{nm}$ summation. The function $\tilde{F}(E_{n},E_{m})$ can be further approximated by 
\begin{eqnarray}
\tilde{F}(E_{n},E_{m})&=&\left(\pi\frac{\partial f(E_{n})}{\partial E_{n}}\right)\delta_{\eta}(E_{n}-E_{m})=\int d\omega\,\delta(\omega-E_{n})\left(\pi\frac{\partial f(\omega)}{\partial \omega}\right)\delta_{\eta}(\omega-E_{m})
\nonumber \\
&\approx&\int d\omega\,\frac{\eta}{(\omega-E_{n})^{2}+\eta^{2}}\left(\frac{1}{\pi}\frac{\partial f(\omega)}{\partial \omega}\right)\frac{\eta}{(\omega-E_{m})^{2}+\eta^{2}}\;,
\end{eqnarray}
\end{widetext}
which leads to Eq.~(\ref{chi_Fnm}). In addition, the vanishing diagonal elements $\tilde{F}(E_{n},E_{n})=0$ are imposed according to Eq.~(\ref{piij_operator}).

Although linear response theory of this kind has been widely adopted to investigate metallic systems, we uncover a number of numerical subtleties that must be implemented for the BHZ model. Firstly, for a homogeneous isolated BHZ model, all states are doubly degenerate due to Kramers' degeneracy (see the block-diagonal form of Eq.~(\ref{BHZ_Dirac_Hamiltonian})). In addition, the edge state at the $y=1$ edge and that at the $y=N_{y,TI}$ edge at the same energy are degenerate, and hence the numerically observed wave function can be an arbitrary mixture of them, which complicates the calculation of the $\langle n|\sigma^{b}(i)|m\rangle$ and $\langle m|\sum_{j}J^{0}(j)|n\rangle$ matrix elements in Eq.~(\ref{chi_Fnm}). Moreover, we find that for a homogeneous BHZ model, the matrix elements $\langle m|\sum_{j}J^{0}(j)|n\rangle$ between the edge states are zero. This is because this matrix element is essentially that of the velocity operator, of which the edge states are eigenstates, so $\langle m|\sum_{j}J^{0}(j)|n\rangle\propto \langle m|{\hat{v}}_{x}|n\rangle=v_{F}\langle m|n\rangle=0$ if $|n\rangle\neq|m\rangle$.

To remove these numerical ambiguities, we focus on the BHZ/FMM junction case in the presence of disorder and a small edge magnetic field for the following reasons. The first advantage of the BHZ/FMM junction is that the degeneracy between the two edges of the BHZ region is lifted due to the nonzero interface coupling $t_{B}\neq 0$ to the FMM at $y=N_{y,TI}$, hence the problem of mixing the wave functions of the two edges is resolved. On the other hand, the $y=1$ edge still accurately captures the Edelstein effect of an isolated BHZ model, so it can be used to compare with the $t_{B}\neq 0$ cases at the other edge $y=N_{y,TI}$. Secondly, the edge state in the presence of disorder is no longer an eigenstate of the velocity operator, so the vanishing $\langle m|\sum_{j}J^{0}(j)|n\rangle$ is resolved. The disorder also helps to smear out the sparce edge state energy spectrum, which increases the accuracy of the numerical calculation. We also add a small magnetic field ${\bf B}(y=1)=B_{y=1}{\hat{\bf z}}$ at the free edge $y=1$ to lift the degeneracy between the two spins, and choose $|n-m|>1$ to avoid neighboring energy levels, such that there is no ambiguity in calculating the  matrix element of the spin operator $\langle n|\sigma^{b}(i)|m\rangle$. Numerically, we perform the calculation on a lattice of size $N_{x}\times(N_{y,TI}+N_{y,FM})=48\times(10+6)$ with $10\%$ impurities of impurity potential $U_{imp}=4$, and use the artificial broadening $\eta=0.1$ (mean free time $\tau\sim 10^{-14}$s) and temperature $k_{B}T=0.03$. Because the function $\tilde{F}(E_{n},E_{m})$ highly peaks at the chemical potential $E_{n}\approx E_{m}\approx 0$, the summation $\sum_{n,m}$ in Eq.~(\ref{chi_Fnm}) is over the $100$ states nearest to the chemical potential. Such a calculation can achieve about $70\%\sim 80\%$ accuracy, which is sufficient to draw conclusions. The accuracy can certainly be improved at larger system sizes.

%The numerical values of $\chi^{b}(i)$ increases with the scattering rate $\eta=\hbar/\tau$, as expected. Using a typical value $\eta=0.05t$ (mean free time $\tau\sim 5\times 10^{-14}$s), we obtain a 

The obtained numerical value of $\chi^{b}$ is of the order of ${\cal O}(1)\times ae/t\sim 10^{-9}$mC/J. Given the typical external charge current in experiment $j_{c}\sim 10^{11}$A/m$^{2}$ and the electrical conductivity of the FMM $\sim 10^{7}$S/m, the corresponding electric field is $E\sim 10^{4}$kgm/C$s^{2}$, which yields a spin polarization $\sigma^{b}(i)\sim 10^{-5}$. Using $J_{ex}=0.1$eV, the spin torque at this typical current density is essentially the numerical values of $\chi^{b}(i)$ averaged over the FMM sites and then multiplied by GHz.

\bibliography{Literatur}

\end{document}